\documentclass[preprint,aps]{revtex4}
\usepackage{mathrsfs}
\usepackage{graphicx}
\usepackage{multirow}
\usepackage{makecell}
\usepackage{booktabs}
\usepackage{bm}

\begin{document}

\title{Spectroscopic Evidence on Realization of a Genuine Topological Nodal Line Semimetal in LaSbTe}

\author{Yang Wang$^{1,2\sharp}$, Yuting Qian$^{1,2\sharp}$, Meng Yang$^{1,2.3\sharp}$, Hongxiang Chen$^{1,4\sharp}$, Cong Li$^{1,2\sharp}$, Zhiyun Tan$^{5}$, Yongqing Cai$^{1,2}$, Wenjuan Zhao$^{1,2}$, Shunye Gao$^{1,2}$, Ya Feng$^{6}$, Shiv Kumar$^{7}$, Eike F. Schwier$^{7,8}$, Lin Zhao$^{1}$, Hongming Weng$^{1,2,9,10}$, Youguo Shi$^{1,2,3}$, Gang Wang$^{1,2,9}$, Youting Song$^{1}$, Yaobo Huang$^{11}$, Kenya Shimada$^{7}$, Zuyan Xu$^{12}$, X. J. Zhou$^{1,2,6,9*}$ and Guodong Liu$^{1,9*}$
}

\affiliation{
\\$^{1}$Beijing National Laboratory for Condensed Matter Physics, Institute of Physics, Chinese Academy of Sciences, Beijing 100190, China
\\$^{2}$University of Chinese Academy of Sciences, Beijing 100049, China
\\$^{3}$Center of Materials Science and Optoelectronics Engineering, University of Chinese Academy of Sciences, Beijing 100049, China 
\\$^{4}$School of Materials Science and Engineering, Fujian University of Technology, Fuzhou 350118, China
\\$^{5}$School of Physics and Electronic Science, Zunyi Normal College, Zunyi 563006, China
\\$^{6}$Beijing Academy of Quantum Information Sciences, Beijing 100193, China
\\$^{7}$Hiroshima Synchrotron Radiation Center, Hiroshima University, Higashi-Hiroshima, Hiroshima 739-0046, Japan
\\$^{8}$Experimentelle Physik VII, Universit$\ddot{a}$t W$\ddot{u}$rzburg, Am Hubland, D-97074 W$\ddot{u}$rzburg, Germany, EU
\\$^{9}$Songshan Lake Materials Laboratory, Dongguan 523808, China
\\$^{10}$CAS Center for Excellence in Topological Quantum Computation, University of Chinese Academy of Science, Beijing 100190, China
\\$^{11}$Shanghai Synchrotron Radiation Facility, Shanghai Advanced Research Institute, Chinese Academy of Sciences, Shanghai 201204, China
\\$^{12}$Technical Institute of Physics and Chemistry, Chinese Academy of Sciences, Beijing 100190, China
\\$^{\sharp}$These people contributed equally to the present work.
\\$^{*}$Corresponding authors: XJZhou@iphy.ac.cn and gdliu$\_$ARPES@iphy.ac.cn
}

\date{February. 22, 2021}


\begin{abstract}
The nodal line semimetals have attracted much attention due to their unique topological electronic structure and exotic physical properties. A genuine nodal line semimetal is qualified by the presence of Dirac nodes along a line in the momentum space that are protected against the spin-orbit coupling. In addition, it requires that the Dirac points lie close to the Fermi level allowing to dictate the macroscopic physical properties. Although the material realization of nodal line semimetals have been theoretically predicted in numerous compounds, only a few of them have been experimentally verified and the realization of a genuine nodal line semimetal is particularly rare. Here we report the realization of a genuine nodal line semimetal in LaSbTe. We investigated the electronic structure of LaSbTe by band structure calculations and angle-resolved photoemission (ARPES) measurements. Taking spin-orbit coupling into account, our band structure calculations predict that a nodal line is formed in the boundary surface of the Brillouin zone which is robust and lies close to the Fermi level. The Dirac nodes along the X-R line in momentum space are directly observed in our ARPES measurements and the energies of these Dirac nodes are all close to the Fermi level. These results constitute clear evidence that LaSbTe is a genuine nodal line semimetal, providing a new platform to explore for novel phenomena and possible applications associated with the nodal line semimetals.
\end{abstract}

\maketitle

Stimulated by the discovery of topological insulators \cite{Hasan_PRL,X.L.Qi_Rev. Mod. Phys,H.Zhang_Nat.Phys,L.Fu_PRB, D.Hsieh_Nature,Y.L.Chen_Science2009,Y.Xia_NP2009_398}, searching for new types of topological materials with novel quantum states has become one of the central topics in current condensed-matter physics \cite{C.L.Kane_PRL,H.M.Weng_MRS Bull,N.P.Armitage_Rev. Mod. Phys,ZhangT_Nat2019_475,VergnioryMG_Nat2019_566,TangF_Nat2019_566, YuanfengXu_Nat2020}. Recently, topological semimetals have attracted intensive interest for their exotic physical properties and potential applications \cite{ZKLiu_Science2014_343,T.Liang_NatMater2015_280,Z.K.Liu_Nat.Mater2014,SonDT_PRB2013_104412,VazifehMM_PRL2013_027201,OjanenT_PRB2013_245112,JXiong_Science2015_413}. The band crossings of a topological semimetal can be categorized by their dimensions in momentum space: the zero-dimensional nodal point \cite{Z.Wang_PRB, H.M.Weng_PRB, H.M.Weng_PRX}, one-dimensional nodal line \cite{Burkov_PRB2011, C.Fang_CPB}, and two-dimensional nodal surface \cite{QFL_PRB, CZhong_Nanos, WWu_PRB}. For the first case, when two or more bands cross each other at a discrete point in Brillouin zone (BZ), it can form Dirac point \cite{Z.K.Liu_Nat.Mater2014, Z.Wang_PRB, H.M.Weng_PRB,ZKLiu_Science2014_343, Neupane_NC2014}, Weyl point \cite{X.Wan_PRB, H.M.Weng_PRX, B.Q.Lv_NP, J.Jiang_NC, CLWang_PRB2016, EHaubold_NM, Hasan_NC, Z.K.Liu_Nat.Mater}, threefold \cite{HMWeng_PRB2016_241202, Z.Zhu_PRX, BQLv_Nature} or even multifold nodal point \cite{Z.P.Sun_PRB2020_155114}. Different from nodal points, one-dimensional nodal curve can take various shapes such as a nodal line, a nodal ring forming a closed loop, or a nodal chain consisting of several inter-connected loops \cite{C.Fang_CPB}. The material realization of the nodal line semimetals have been theoretically predicted in many compounds \cite{Burkov_PRB2011, Yang_Adv, ZFang_FrontPhys, QXu_HMW_PRB2017_045136, MYang_PRM2020}, but only a few of them have been experimentally examined by angle-resolved photoemission spectroscopy (ARPES) such as PtSn$_4$ \cite{YWu_NP}, CaAgX (X=P, As) \cite{Takane_npj2018_1, CLiu_PRB,Xu_PRB2018_161111}, Ta$_3$SiTe$_6$ \cite{TS_PRB2018_121111}, monolayer Cu$_2$Si \cite{BJFeng2017IMatsuda_NC}, XB$_2$ (X=Zr, Ti, Al, Mg) \cite{RLuo2018SCWang_npjQM, ZHLiu2018SCWang_PRX, CJYi2018HDing_PRB, DTakane2018TSato_PRB, XQZhou2019DSDessau_PRB}, RAs$_3$ (R=Ca,Sr) \cite{MMHosen2020MNeupane_SP, YKSong2020DWShen_PRL}, Pt$_2$HgSe$_3$ \cite{ICucchi2020ATamai_PRL}, Co$_2$MnGa \cite{HBelopolski2019MZHasan_S}, ZrSiS family \cite{Madhab Neupane_PRB, XFWang2017RZhang_AM, T.Takahashi_PRB, Schoop_NC, Leslie M. Schoop_New J. Phys, Madhab_PRB2017, CChen_PRB2017_125126, YunYen_arxiv, Fu_Sci.Adv, Hosen_ScienReport, MMHosen2018MNeupane_PRB, BChen_SCPMA2020, SYue_BJFeng_PRB2020}, single-layer GdAg$_2$ \cite{BaoJieFeng_PRL2019_116401}, InBi \cite{SAEkahana2017YLChen_NJP}, PbTaSe$_2$ \cite{G.Bian_PRB2018}, IrO$_2$ \cite{Nelson2019Moreschini_PRM} and RuO$_2$ \cite{Jovic2019Moser_arxiv}.

Strictly speaking, two requirements are necessary for realizing a genuine nodal line semimetal. One is the presence of Dirac nodes along a line and the nodes retain gapless under the spin-orbit coupling (SOC). The other is that the Dirac points should lie close to the Fermi level allowing to produce exotic physical phenomena. Many of the nodal line semimetals proposed without considering SOC may not fall into the category because the Dirac points become unstable and get gapped when SOC is considered, as seen in CaAgX (X=P, As) \cite{Takane_npj2018_1, CLiu_PRB,Xu_PRB2018_161111}, Ta$_3$SiTe$_6$ \cite{TS_PRB2018_121111}, monolayer Cu$_2$Si \cite{BJFeng2017IMatsuda_NC}, XB$_2$ (X=Zr, Ti, Al, Mg )\cite{RLuo2018SCWang_npjQM, ZHLiu2018SCWang_PRX, CJYi2018HDing_PRB, DTakane2018TSato_PRB, XQZhou2019DSDessau_PRB}, RAs$_3$ (R=Ca, Sr) \cite{MMHosen2020MNeupane_SP, YKSong2020DWShen_PRL}, Pt$_2$HgSe$_3$ \cite{ICucchi2020ATamai_PRL} and Co$_2$MnGa \cite{HBelopolski2019MZHasan_S}. The genuine topological nodal line semimetals that have been proposed and can satisfy the two criteria are rare \cite{Nelson2019Moreschini_PRM, Jovic2019Moser_arxiv}. The layered ternary WHM system (W represents transition metal Zr, Hf, or rare earth element La, Ce, Gd; H represents group IV or group V element Si, Ge, Sn, or Sb, and M represents group VI element O, S, Se, and Te) \cite{XDai_PRB} has become an important candidate family of topological nodal line semimetals. Without SOC or when the SOC is negligible, this system can host nodal lines and nodal surface, which have been confirmed experimentally by ARPES measurements in HfSiS \cite{T.Takahashi_PRB}, ZrSiX (X=S, Se, Te) \cite{Madhab Neupane_PRB, XFWang2017RZhang_AM, Schoop_NC, Madhab_PRB2017, Fu_Sci.Adv}, ZrGeTe \cite{MMHosen2018MNeupane_PRB,YunYen_arxiv} and GdSbTe \cite{Hosen_ScienReport}. When the SOC is taken into account, however, most of the nodal lines and nodal surface become gapped, leaving only two topological nodal lines in the boundary surface of Brillouin zone \cite{CChen_PRB2017_125126, BChen_SCPMA2020}. Meanwhile, the robust nodal lines observed so far in this system stay away from the Fermi level, contributing little to the macroscopic physical properties. Finding genuine topological nodal line semimetals is essential for exploring new phenomena and realizing exotic properties. 

In this paper, we report identification of LaSbTe as a genuine topological nodal line semimetal. Compared to the ZrSiS system in which the robust Dirac nodal lines exist at the Brillouin zone boundary but their energy position stays away from the Fermi level ($\sim$0.6 eV below the Fermi level in ZrSiS and HfSiS\cite{CChen_PRB2017_125126}), LaSbTe  possesses not only the robust nodal lines but also these nodal lines locate right  near the Fermi level so that the exotic topological properties may manifest.  We carried out band structure calculations and ARPES measurements to investigate the topological electronic structure of LaSbTe. Our band structure calculations combined with symmetry analyses indicate that two nodal lines along X-R and M-A lines in momentum space can be formed and stay robust after the SOC is considered. Among these two nodal lines, the one along the X-R direction lies close to the Fermi level while the other one along M-A direction stays far from the Fermi level. Our ARPES measured results are in good agreements with the band structure calculations. In particular, we clearly observed Dirac nodes along the X-R line with the Dirac points close to the Fermi level. These results provide clear evidence that LaSbTe is a genuine topological nodal line semimetal.

High-quality single crystals of LaSbTe were synthesized by flux method. The samples are characterized by single crystal X-ray diffraction measurements. Single crystal X-ray diffraction measurement was conducted on Bruker D8 high resolution four-circle diffractometer at 273 K using Mo K$\alpha$ radiation ($\lambda$=0.71073 $\AA$), indicating that our LaSbTe samples crystallize in a nonsymmorphic space group $P4/nmm$ (no.129) with PbFCl-type tetragonal structure: a=b=4.399 $\AA$ and c=9.566 $\AA$. The measured crystal
parameters of LaSbTe are summarized in Table I-III in Supplementary Materials\cite{SM}. This structure is identical to the WHM family. The Sb atoms constitute a two-dimensional square net sandwiched in between two LaTe layers (Fig. 1a). ARPES measurements were performed at Hiroshima Synchrotron Radiation (HiSOR) BL-1 Beamline \cite{Shimada_NIMA2001,Iwasawa_JSR2017}, the `Dreamline' Beamline of the Shanghai Synchrotron Radiation Facility (SSRF) and the ARPES system in our lab \cite{Liu_RSI2008, Zhou_RPP2018}. The energy resolution is $\sim$30 meV and the angular resolution is $\sim$0.3 degree. In order to get complete electron structure, two polarization geometries were used in our ARPES measurements. In one polarization geometry (LV), the electric field vector of the incident light is perpendicular to the horizontal plane, while in the other polarization geometry (LH), it lies within the horizontal plane. All the samples were cleaved {\it in situ} and measured at different photon energies in ultrahigh vacuum with a base pressure better than 5.0$\times$10$^{-11}$ mbar. The first-principles band-structure calculations based on the density functional theory (DFT) have been done using the Vienna ab initio simulation package (VASP) \cite{Georg_PRB1996_11169} within the generalized gradient approximation (GGA) of Perdew-Burke-Ernzerhof type \cite{John_PRL1996_3868}. We take 300 eV as the cut-off energy for plane wave expansion. The 8 $\times$ 8 $\times$ 4 k-mesh is used as grids of Brillouin zone (BZ) in the self-consistent process. The relaxed lattice parameters a=b=4.421 $\AA$, c=9.659 $\AA$ are employed. The Wannier90 \cite{Nicola_RMP2012_1419} and WannierTools \cite{QuanSheng Wu_CPC2018_416} are used to calculate the Fermi surface and find nodes based on the maximally localized Wannier functions which consist of p orbits of Sb and Te, and d orbits of La.

 \hspace*{\fill}

We start our discussion with bulk electronic structure calculations of LaSbTe. Symmetries are crucial to understand the electronic structure of materials as they protect the band crossings in the Brillouin zone in high-symmetry planes, lines or points\cite{C.Fang_CPB, WWu_PRB}. For the tetragonal LaSbTe crystallized in nonsymmorphic space group $P4/nmm$, the glide mirror $\tilde{\mathcal{M}}_{z}=\left\{ \mathcal{M}_{z}|\frac{1}{2} \frac{1}{2} 0 \right\}$, the mirror $\tilde{\mathcal{M}}_{x}=\left\{ \mathcal{M}_{x}|\frac{1}{2} 0 0 \right\}$, $\tilde{\mathcal{M}}_{y}=\left\{ \mathcal{M}_{y}|0 \frac{1}{2} 0 \right\}$, $\tilde{\mathcal{M}}_{xy}=\left\{ \mathcal{M}_{xy}|\frac{1}{2} \frac{1}{2} 0 \right\}$, the spacial inversion $\left\{ \mathcal{P}|0 0 0 \right\}$ symmetries and the two screw symmetries: $\tilde{\mathcal{C}}_{2x}=\left\{ \mathcal{C}_{2x}|\frac{1}{2} 0 0 \right\}$ and  $\tilde{\mathcal{C}}_{2y}=\left\{ \mathcal{C}_{2y}|0 \frac{1}{2} 0 \right\}$ are especially important. Combining with time-reversal symmetry $\mathcal{T}$, these symmetries can lead to nonsymmorphically enforced degeneracy at the boundary of Brillouin zone that give rise to multiple nodal lines. To clarify the topology of electronic structure of LaSbTe protected by different symmetries, we calculated the Dirac nodal lines in three-dimensional (3D) Brillouin zone (Fig. 1c) and the band structures of bulk LaSbTe without (Fig. 1d) and with SOC (Fig. 1e). Because the bands close to the Fermi level are mainly dominated by the $\alpha$, $\beta$ and $\gamma$ in Fig. 1d and Fig. 1e, we will focus on the nodal lines formed by these three bands. 

Figure 1c shows the schematic of the calculated nodal line configurations formed by $\beta$-$\gamma$ band (NL1-NL5) in the absence of SOC and by $\alpha$-$\beta$ band (NLE1 and NLE2) in the presence of SOC in 3D Brillouin zone. They represent the locations of all the Dirac nodes in three-dimensional momentum space formed from $\alpha$, $\beta$ and $\gamma$ bands although the energy of the Dirac nodes can be different. The nodal lines formed by $\beta$-$\gamma$ band can take various shapes under different symmetry protections as shown by the light blue curves in Fig. 1c. They all lie inside the first Brillouin zone. We can find two diamond-shaped nodal lines: NL1 in k$_{z}$=0 plane and NL2 in k$_{z}$=$\pi$ plane that are protected by the glide mirror symmetry $\tilde{\mathcal{M}}_{z}$. In addition, there are two vertical nodal lines, NL3 protected by $\tilde{\mathcal{M}}_{x}$ and NL4 by $\tilde{\mathcal{M}}_{xy}$. These nodal lines form a 3D cage-like structure. They are 2-fold degenerate and not stable in the presence of SOC which will open a gap at the Dirac point.

The band crossings formed by $\alpha$-$\beta$ band form two nodal surfaces in the k$_x$=$\pi$ and k$_y$=$\pi$ planes due to the combined symmetry $\tilde{\mathcal{C}}_{2y}\mathcal{T}$ as reported in WHM family \cite{Fu_Sci.Adv,LMuechler_PRX2019}. When SOC is taken into consideration, all band crossings are gapped except for the two lines along R-X and A-M directions and some one-dimensional curves at generic k points in A-M-X-R-A plane owing to accidental degeneracy that are far away from Fermi level and will not be considered further. The nodal lines along R-X and A-M high symmetry directions, as labeled by NLE1 and NLE2 in Fig. 1c, host robust 4-fold degenerate massless Dirac nodal lines protected by nonsymmorphic symmetry at the boundary of Brillouin zone \cite{C.Fang_CPB, CChen_PRB2017_125126, Schoop_NC, SMYoung_PRL2015, CFang_PRB2015_081202} and can be understood in the following way. Firstly, in the spinful case, it is well known that there exists Kramers' pair due to $(\mathcal{PT})^2$=-1. Secondly, the two high symmetry lines R-X and A-M have higher symmetry than other points on the k$_y$=$\pi$ plane such that bands along these two lines are at least doubly degenerate ($|\psi_1\rangle, |\psi_2\rangle$) protected by $\tilde{\mathcal{M}}_y$. Thirdly, since the $\tilde{\mathcal{M}}_y$ is a nonsymmorphic symmetry, anti-commutation relation $\{\mathcal{PT},\tilde{\mathcal{M}}_y\}$=0 is satisfied. Considering the above three conditions, there must exist a 4-fold degenerate state ($|\psi_1\rangle, \mathcal{PT}|\psi_1\rangle, |\psi_2\rangle, \mathcal{PT}|\psi_2\rangle$) where each Kramers' pair ($|\psi_{1,2}\rangle, \mathcal{PT}|\psi_{1,2}\rangle$) belongs to two orthogonal states with the same $\tilde{\mathcal{M}}_y$ eigenvalues \cite{C.Fang_CPB, CFang_PRB2015_081202}. Therefore, the NLE1 and NLE2 are symmetry enforced and appear along high symmetry lines at the boundary of Brillouin zone.

The formation of nodal lines in LaSbTe can be seen directly from the calculated band structures  which show the overall $\alpha$, $\beta$ and $\gamma$ bands, derived mainly from the Sb p$_x$ and p$_y$ orbits, in the entire 3D Brillouin zone. Without considering SOC, the location of the Dirac points or nodal lines formed from $\alpha$-$\beta$ band (red arrows) and $\beta$-$\gamma$ band (dashed black boxes) are marked in Fig. 1d. When SOC is taken into consideration as shown in Fig. 1e, all the Dirac points formed by $\beta$-$\gamma$ band are gapped. But some Dirac points formed by $\alpha$-$\beta$ band remain intact. The SOC causes the band splitting along M-X (A-R) direction and turns the initial Dirac line nodes (marked by black arrows in Fig. 1d) into topologically trivial 2-fold spin-degenerate bands. However, the two Dirac lines along the R-X and A-M lines survive under the protection of nonsymmorphic symmetry, as labeled by red arrows in Fig. 1e. The NLE2 nodal line lies about 1.6 eV above the Fermi level while the NLE1 nodal line lies right at the Fermi level. These calculation results indicate that LaSbTe is a genuine nodal line semimetal.

Now we come to the electronic structure of LaSbTe from our ARPES measurements. Fig. 2a-2c show the Fermi surface mapping and constant energy contours of LaSbTe measured with photon energies of 55 eV (Fig. 2a), 85 eV (Fig. 2b) and 95 eV (Fig. 2c). We can clearly see a large diamond-shaped Fermi surface centered around $\bar\Gamma$ point. We also observe strong spectral weight at the $\bar{X}$ points. For different photon energies, the evolution of constant energy contours with binding energy is markedly different for the three different photon energies. Fig. 2d-2f show the calculated Fermi surface and constant energy contours of LaSbTe at k$_{z}$ planes of 0.5 $\pi/c$ (Fig. 2d), 0.75 $\pi/c$ (Fig. 2e) and 0 $\pi/c$ (Fig. 2f), corresponding to the photon energy of 55 eV, 85 eV and 95 eV, respectively. The diamond-shaped Fermi surface originates from $\beta$ band and the feature around $\bar{X}$ point is mainly from the Dirac-like band formed from $\alpha$ or $\beta$ bands, as shown in Fig. 1e. The measured results show an overall agreements with the band structure calculations in terms of the observation of a large diamond-shaped Fermi surface and the feature at the $\bar{X}$ points. 
 
Figure 3 shows the band structure of LaSbTe measured along three high symmetry directions under different polarization geometries (more data measured at different photon energies can be found in Supplementary Fig. S2 and Fig. S3\cite{SM}). We find that the observed band structures measured under different polarizations are quite different due to photoemission matrix element effect.  Therefore, our measurements under two distinct polarizations are helpful to directly reveal the band structure depending on the orbital symmetry with respect to the detection plane. To better identify the band structure, we also take second derivative images (Fig. 3d-3f) of the original data (Fig. 3a-3c). For comparison, we put the corresponding calculated bulk band in Fig. 3g-3i and carried out slab calculations (Fig. 3j-3l) that can handle surface states.

For the band structure along $\bar\Gamma$-$\bar{M}$ direction, a linearly dispersive band crosses the Fermi level as shown in Fig. 3d marked by red and orange lines, which gives rise to the diamond-shaped Fermi surface. This band can extend to a high binding energy of $\sim$2 eV. It has a very high Fermi velocity of $\sim$12 eV$\cdot{\AA}^{-1}$ that is larger than those found in some typical Dirac materials like graphene (6.7 eV$\cdot{\AA}^{-1}$) \cite{WeiYao_PRB2015_115421} , SrMnBi$_2$ (10.6 eV$\cdot{\AA}^{-1}$) \cite{YaFeng_ScientificReports2014_5385} and ZrSiS (4.3 eV$\cdot{\AA}^{-1}$) \cite{Schoop_NC}. The measured band structures (Fig. 3d) areconsistent with the calculated bands (Fig. 3g) in terms of both the number and the position of the bands. For the band structure along $\bar\Gamma$-$\bar{X}$ direction, a Dirac like band is observed crossing the Fermi level as shown in Fig. 3e marked by red and orange lines. The Dirac point lies exactly at $\bar{X}$ point and its energy position is rather close to the Fermi level. Again, the measured band structures (Fig. 3e) are consistent with the calculation (Fig. 3h) in terms of the number, shape and the position of the bands. For the band structure along $\bar{M}$-$\bar{X}$ direction, a tiny electron-like band is observed that crosses the Fermi level at the $\bar{X}$ point as shown in Fig. 3f. In addition, two groups of bands are observed, labeled as SB and BB in Fig. 3f. Each group of band is further composed of two bands. These results are well reproduced in the band structure calculations (Fig. 3i and 3l), which indicate that the SB bands represent surface states while the BB bands are bulk states. The SB band exhibits little change when measured at different photon energies (Fig. S3 in Supplementary materials\cite{SM}), further confirming its surface state nature.

It is noted that we only observed single sheet of the diamond-shaped Fermi surface, even at different binding energies (Fig. 2a-2c). According to the band structure calculations, two sheets of the Fermi surface are expected (Fig. 2d-2f). We also only observed a single band over a large energy range ($\sim$0.6 eV) near the Fermi level measured along $\bar\Gamma$-$\bar{M}$ direction (Fig. 3a and Fig. 3d) although two linear bands are expected in this energy range from the band calculations (Fig. 1d and Fig. 1e). Most of the  ARPES measurements on WHM system have observed double sheets Fermi surface \cite{T.Takahashi_PRB, Madhab Neupane_PRB, Leslie M. Schoop_New J. Phys, Madhab_PRB2017, CChen_PRB2017_125126, YunYen_arxiv, Fu_Sci.Adv}. A single sheet Fermi surface has also been reported before in GdSbTe but the origin is unclear \cite{Hosen_ScienReport}. By carefully examing the measured band structures in Fig. 3d, we believe this is caused by the photoemission matrix element effect. As seen in the right panel of Fig. 3d, in this particular polarization geometry, the two linear bands can actually be observed at high binding energy between 1.0$\sim$1.5 eV although the right-side band is weaker than the left-side one. When moving to lower binding energy between 0$\sim$1.0 eV, the right-side band gets strongly suppressed while the left-side one remains strong and becomes dominant. When the sample is measured in another polarization geometry (left panel of Fig. 3d), the right-side band is fully suppressed. The spectral weight suppression of the right-side linear band results in our observation of single sheet diamond-like Fermi surface and a single band near the Fermi level. 

From the band structure calculations, the Dirac nodes along the diamond-shaped Fermi surface are gapped when the SOC is considered (Fig. 1d and 1e). The calculated gap size is about 40 meV. To check on the gap opening, we carefully examined the band structures measured along $\bar\Gamma$-$\bar{M}$ direction using different photon energies (Fig. S2 in Supplementary materials\cite{SM}). We observed sharp single band over an energy range of 0$\sim$0.6 eV below the Fermi level. There is a finite spectral intensity at the Fermi level and the band intensity shows a smooth variation with the binding energy (Fig. S2i-S2l). We do not observe clear signatures of gap opening along the $\bar\Gamma$-$\bar{M}$ direction. This is probably because the gap opening induced by SOC is small that exceeds our present detection limit. 

As predicted by band structure calculations, LaSbTe is a nodal line semimetal because of the presence of the nodal line along X-R direcion that is robust against the SOC and lies close to the Fermi level. To examine on the band structure along k$_z$ direction, we carried out measurements along $\bar\Gamma$-$\bar{X}$ direction with different photon energies between 26 eV and 95 eV. Fig. 4a shows the Fermi surface mapping in $\Gamma$-X-R-Z-$\Gamma$ plane of LaSbTe crossing the $\bar{X}$ point along $\bar{\Gamma}$-$\bar{X}$ direction. A one-dimensional linear Fermi surface is clearly observed along the X-R momentum line. Fig. 4b displays representative band structures measured using different photon energies between 32 eV and 50 eV, that corresponds to a full k$_z$ period. Dirac-like structure can be observed in all the measurements using different photon energies as marked by the red arrows in Fig. 4b. The Dirac cones lie at the $\bar{X}$ point along the X-R line in momentum space and their energies are all close to the Fermi level. These results strongly indicate that there is a nodal line formed along the X-R direction.

In summary, by carrying out ARPES measurements combined with band structure calculations, we have systematically investigated the electronic structures of LaSbTe. Our band structure calculations predict the formation of five nodal lines derived from the $\beta$-$\gamma$ band including two diamond-like nodal lines, and two nodal surfaces (k$_x$=$\pi$ and k$_y$=$\pi$ plane) from the $\alpha$-$\beta$ band without considering SOC. Taking SOC into account, band calculations indicate that those five nodal lines from $\beta$-$\gamma$ band become unstable and are gapped at the Dirac points while the two nodal lines from $\alpha$-$\beta$ band are robust. Among these two nodal lines, the one along M-A direction stays far away from Fermi level while the other one along  X-R direction lies close to the Fermi level. Our ARPES results are in good agreements with the band structure calculations. We observed a diamond-like Fermi surface. In particular, we directly observed Dirac nodes along the X-R direction with their energy position close to the Fermi level. These results provide strong evidence that LaSbTe is a genuine topological nodal line semimetal, which is the first case that is realized in the WHM family. This finding will facilitate to explore for novel phenomena and possible applications associated with the nodal line semimetals.
\\

\vspace{3mm}

\noindent {\bf Acknowledgement}\\
This work is supported by the National Key Research and Development Program of
China (Nos. 2016YFA0300600, 2018YFA0305602 and 2018YFE0202600), the National Natural Science Foundation of China (Nos. 11974404, U2032204 and 51832010), the Strategic Priority Research Program (B) of the Chinese Academy of Sciences (Nos. XDB33000000 and QYZDB-SSW-SLH043), and the Youth Innovation Promotion Association of CAS (No. 2017013). ARPES measurements at HiSOR were performed under the Proposal Nos. 18AU001 and 17BU025. We thank the N-BARD, Hiroshima University for supplying liquid He. The theoretical calculations is supported by the National Natural Science Foundation of China (Grant Nos.  11674369, 11865019 and 11925408), the Beijing Natural Science Foundation (Grant No. Z180008), Beijing Municipal Science and Technology Commission (Grant No. Z191100007219013), the National Key Research and Development Program of China (Grant Nos. 2016YFA0300600 and 2018YFA0305700), the K. C. Wong Education Foundation (Grant No. GJTD-2018-01) and the Strategic Priority Research Program of Chinese Academy of Sciences (Grant No. XDB33000000).

\vspace{3mm}

\noindent {\bf Author Contributions}\\

Y.W., Y.T.Q., M.Y., H.X.C. and C.L. contributed equally to this work. G.D.L.,  X.J.Z., Y.W. and C.L. proposed and designed the research. M.Y., H.X.C., Y.G.S. and G.W. contributed to LaSbTe crystal growth. Y.T.S. contributed to the single-crystal diffraction measurements and crystal structure analysis. Y.T.Q., Z.Y.T. and H.M.W. contributed to the band structure calculations and theoretical discussion. Y.W. carried out the ARPES experiment at SSRF and HiSOR with the assistance from S.Y.G. (SSRF), Y.B.H. (SSRF), W.J.Z. (HiSOR), Y.F. (HiSOR), S.K. (HiSOR), E.F.S. (HiSOR) and K.S. (HiSOR). W.Y. and C.L. carried out the experiment at home-built ARPES system in our Lab. C.L., Y.Q.C., Y.W.,  G.D.L., L.Z., Z.Y.X. and X.J.Z. contributed to the development and maintenance of our ARPES system. Y.W., C.L., G.D.L. and X.J.Z. analyzed the data. Y.W., C.L., G.D.L. and X.J.Z. wrote the paper with Y.T.Q. and H.M.W.  All authors participated in discussion and comment on the paper.

\newpage

\begin{figure*}[tbp]
\begin{center}
\includegraphics[width=0.8\columnwidth,angle=0]{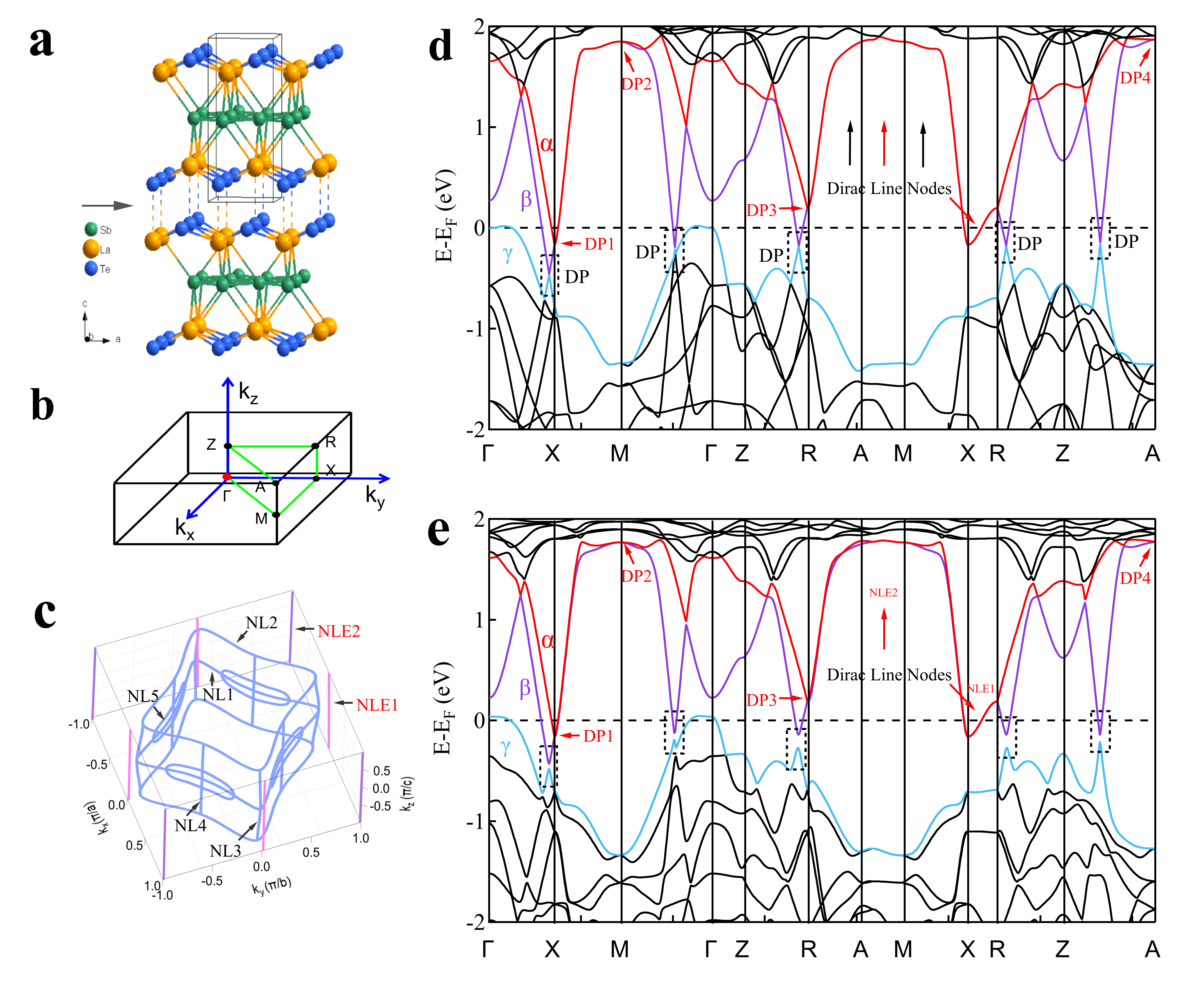}
\end{center}
\caption{\textbf{Band structure calculations and the nodal line configuration of LaSbTe.} (a) Crystal structure of LaSbTe. The Sb atoms shown with blue balls form square nets, which are sandwiched between the two La-Te layers. The grey arrow indicates the cleavage plane between the LaTe layers. (b) Schematic of 3D Brillouin zone of LaSbTe. (c) Calculated nodal lines formed by $\beta$-$\gamma$ band (NL1-NL5) and $\alpha$-$\beta$ band (NLE1 and NLE2) in 3D Brillouin zone. (d-e) Calculated bulk band structures without (d) and with SOC (e) along high symmetry directions by first-principles calculations. We label the two conduction bands and one valence band close to the Fermi level with $\alpha$ (red), $\beta$ (purple) and $\gamma$ (blue)  respectively. The Dirac points are marked by red arrows.
} 
\end{figure*}

\begin{figure*}[tbp]
\begin{center}
\includegraphics[width=1\columnwidth,angle=0]{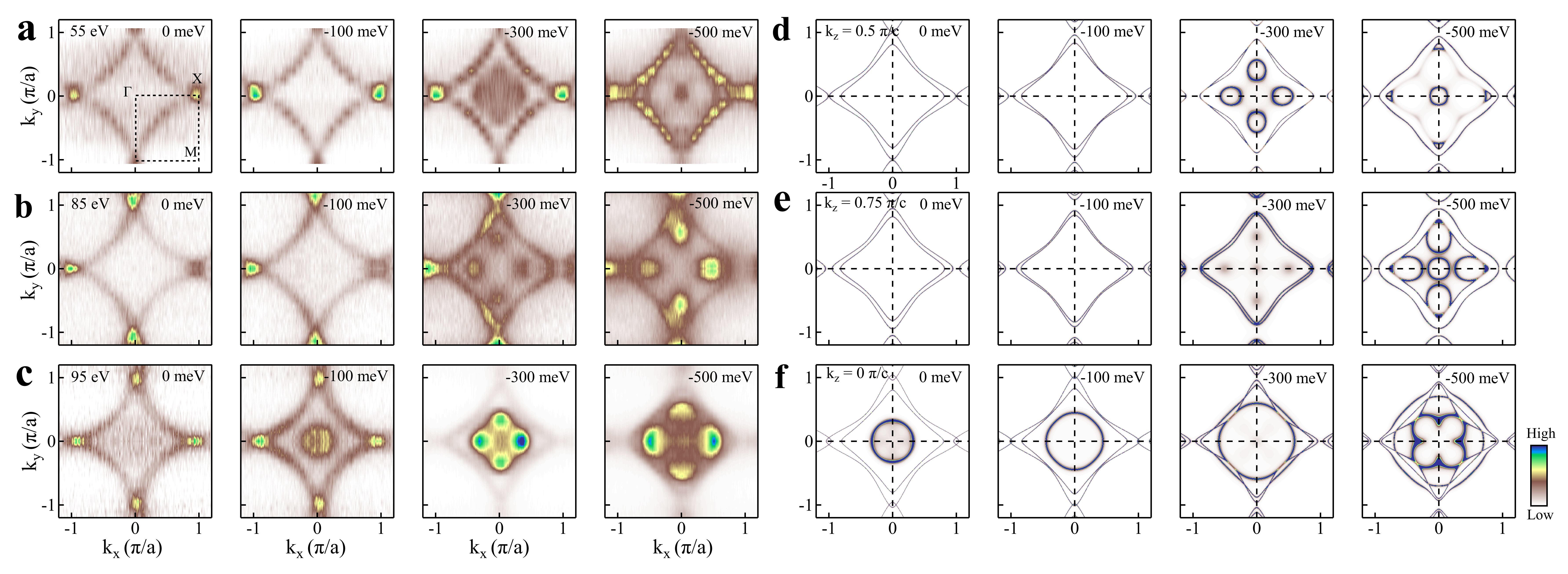}
\end{center}
\caption{\textbf{Fermi surface of LaSbTe.} (a-c) The Fermi surface and constant energy contours of LaSbTe measured with photon energy of 55 eV (a), 85 eV (b) and 95 eV (c), respectively, under $LH$ polarization. (d-f) The DFT calculated Fermi surface and constant energy contours of LaSbTe with k$_z$ corresponding to 0.5 $\pi$/c (d), 0.75 $\pi$/c (e) and 0 $\pi$/c (f), respectively.
}
\end{figure*}

\begin{figure*}[tbp]
\begin{center}
\includegraphics[width=1\columnwidth,angle=0]{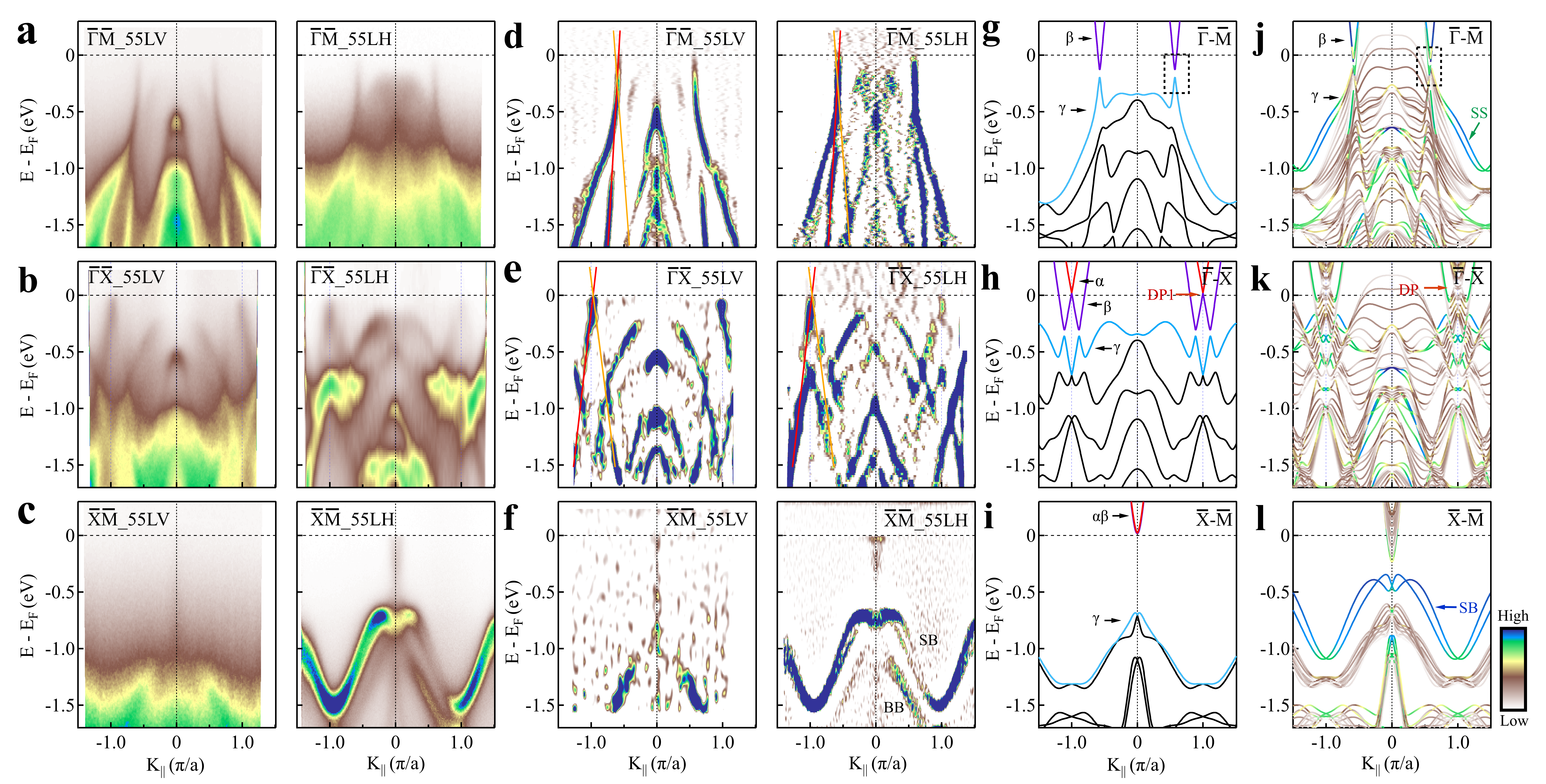}
\end{center}
\caption{\textbf{Measured band structures of LaSbTe and their comparison with band structure calculations.} (a-c) Band structures of LaSbTe measured along $\bar\Gamma$-$\bar{M}$ (a), $\bar\Gamma$-$\bar{X}$ (b) and $\bar{X}$-$\bar{M}$ (c) directions, respectively, with a photon energy of 55 eV. The left and right panels in (a-c) are measured under LV and LH polarization geometries, respectively. (d-f) The second derivative images of the original data corresponding to (a-c).
The red and orange lines mark the two branches of the linearly dispersed Dirac bands. (g-i) The calculated bulk band structures of LaSbTe with SOC along $\bar\Gamma$-$\bar{M}$ (g), $\bar\Gamma$-$\bar{X}$ (h) and $\bar{X}$-$\bar{M}$ (i) momentum directions for k$_z$=0.5$\pi$/c plane. (j-l) The calculated band structures of LaSbTe with SOC along $\bar\Gamma$-$\bar{M}$ (j), $\bar\Gamma$-$\bar{X }$(k) and $\bar{X}$-$\bar{M}$ (l) directions for a seven-unit-cell-thick slab. The intensity of color bar represents the proportion of surface states contributed by the topmost unit cell of LaSbTe. Here, the BS, BB and the DP indicate the bulk band, surface band and the Dirac point, respectively.
}
\end{figure*}

\begin{figure*}[tbp]
\begin{center}
\includegraphics[width=1\columnwidth,angle=0]{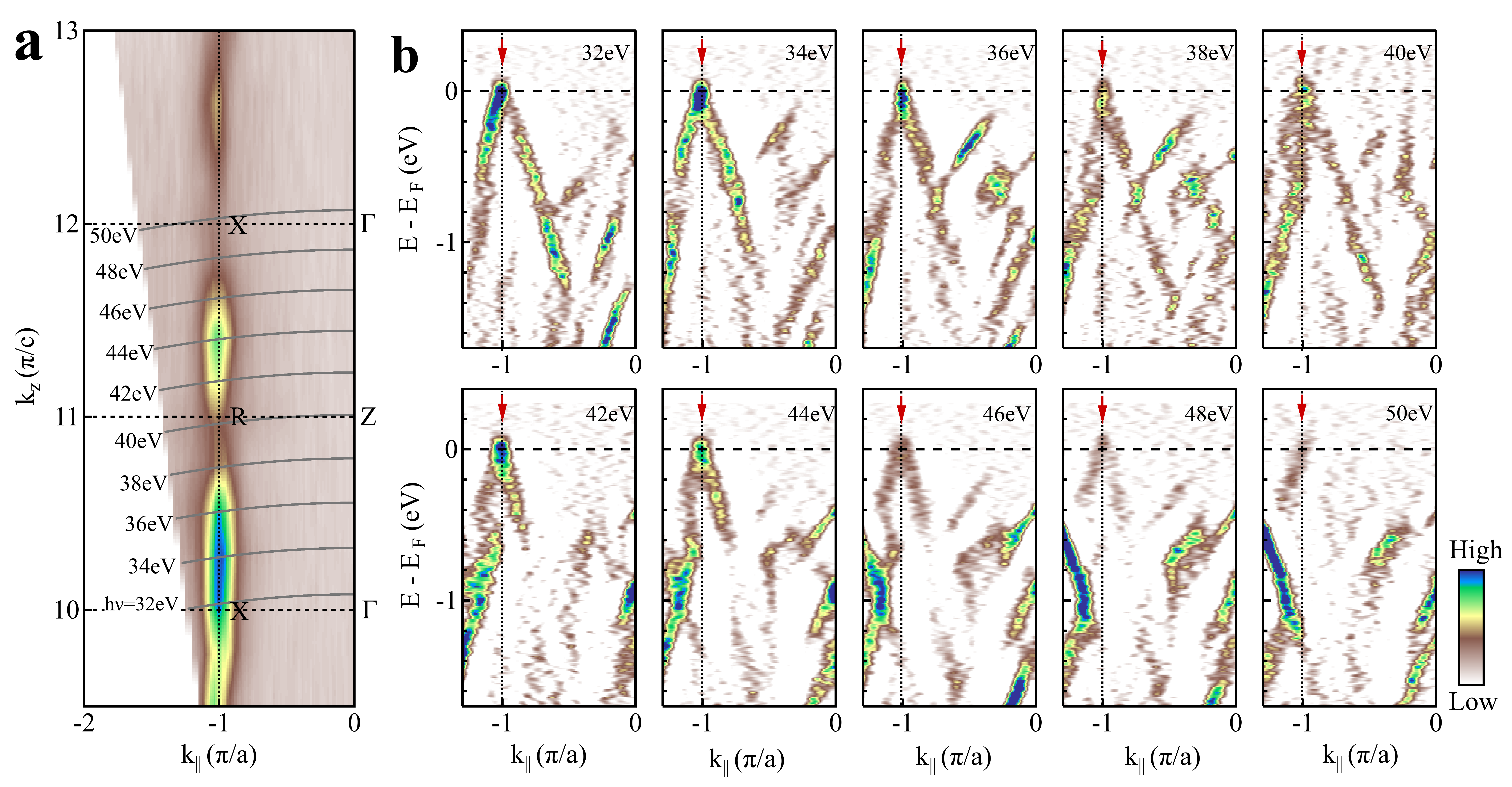}
\end{center}
\caption{\textbf{Dirac nodal line along X-R direction.} (a) Photon energy dependent ARPES-intensity measurement of LaSbTe crossing the zone center along $\bar\Gamma$-$\bar{X}$ direction. (b) Band structures measured along $\bar\Gamma$-$\bar{X}$ direction at different photon energies. To highlight the bands, these images are from the second derivative of the original data. The Dirac points are marked by red arrows.
} 
\end{figure*}

\end{document}